\documentclass[showpacs,preprintnumbers,amsmath,amssymb,nofootinbib]{revtex4}
\usepackage[dvips]{graphicx}% Include figure files
\usepackage{dcolumn}% Align table columns on decimal point
\usepackage{bm}% bold math

\def\mapgeq{\mathbin{\lower.3ex\hbox{$\buildrel>\over{\smash{\scriptstyle\sim}\vphantom{_x}}$}}}
\def\mapleq{\mathbin{\lower.3ex\hbox{$\buildrel<\over{\smash{\scriptstyle\sim}\vphantom{_x}}$}}}
\def\mapgeqeq{\mathbi{\lower.3ex\hbox{$\buildrel>\over{\smash{\scriptstyle\approx}\vphantom{_2}}$}}}
\def\mapleqeq{\mathbin{\lower.3ex\hbox{$\buildrel<\over{\smash{\scriptstyle\approx}\vphantom{_2}}$}}}
\def\Journal#1#2#3#4{{#1} {\bf #2} (#4) #3}
\def\MPL{Mod. Phys. Lett. A}

\def\NPB{Nucl. Phys. B}
\def\NPBOLD{Nucl. Phys.}

\def\NIMA{Nucl. Instrum. Meth A}
\def\PLB{Phys. Lett. B}

\def\PLBOLD{Phys. Lett.}
\def\PRL{Phys. Rev. Lett.}
\def\RMP{Rev. Mod. Phys.}
\def\PRD{Phys. Rev. D}

\def\PTP{Prog. Theor. Phys.}
\def\JHEP{JHEP}
\def\EPJ{Euro. Phys. J. C}
\def\JETPUSSR{JETP (USSR)}
\def\ZETP{Zh. Eksp. Teor. Piz.}

\def\IJMP{Int. J. Mod. Phys. A}

\def\JPG{J. Phys. G}
\def\Erratum{Erratum-ibid}

\begin{document}

%\preprint{TOKAI-HEP/TH-0403}

\title{General Property of Neutrino Mass Matrix and CP Violation\footnote{to appear in Physics Letters B.}}
% Force line breaks with \\

\author{Ichiro Aizawa}
 \email{4aspd001@keyaki.cc.u-tokai.ac.jp}
\author{Masaki Yasu\`{e}}%
\email{yasue@keyaki.cc.u-tokai.ac.jp}
\affiliation{\vspace{5mm}%
\sl Department of Physics, Tokai University,\\
1117 Kitakaname, Hiratsuka, Kanagawa 259-1291, Japan}

\date{October, 2004}% It is always \today, today,

%%-------------------------------------------------
%% Abstruct
%%-------------------------------------------------
\begin{abstract}
It is found that the atmospheric neutrino mixing angle of $\theta_{atm}$ is determined to be $\tan\theta_{atm} = {\rm Im} (B)/{\rm Im} (C)$ for $B$=$M_{\nu_e\nu_\mu}$ and $C$=$M_{\nu_e\nu_\tau}$, where $M_{ij}$ is the $ij$ element of $M^\dagger_\nu M_\nu$ with $M_\nu$ as a complex symmetric neutrino mass matrix in the ($\nu_e$, $\nu_\mu$, $\nu_\tau$)-basis.  Another mixing angle, $\theta_{13}$, defined as $U_{e3} = \sin\theta_{13}e^{-i\delta}$ is subject to the condition: $\tan 2\theta_{13}\propto \vert \sin\theta_{atm} B+\cos\theta_{atm}C\vert$ and the CP-violating Dirac phase of $\delta$ is identical to the phase of $\sin\theta_{atm}B^\ast+\cos\theta_{atm}C^\ast$.  The smallest value of $\vert \sin \theta_{13} \vert$ is achieved at $\tan\theta_{atm} = - {\rm Re}(C)/{\rm Re}(B)$ that yields the maximal CP-violation and that implies $C=-\kappa B^\ast$ for the maximal atmospheric neutrino mixing of $\tan\theta_{atm}=\kappa=\pm 1$.  The generic smallness of $\vert \sin \theta_{13} \vert$ can be ascribed to the tiny violation of the electron number conservation.
\end{abstract}

\maketitle
%%-------------------------------------------------
%% Main body
%%-------------------------------------------------
%%-------------------------------------------------
%% Main body
%%-------------------------------------------------
\section{\label{sec:1}Introduction}
Properties of neutrino oscillations have been extensively studied in various experiments \cite{NeutrinoSummary,RecentData} since the confirmation of the atmospheric neutrino oscillations by the SuperKamiokande collaborations in 1998 \cite{SuperKamiokande}.  It is next expected that leptonic CP-violation can be observed in neutrino-related reactions \cite{CPViolation} via CP-violating phases in the neutrino sector \cite{CPphases}.  The CP-violating phases are introduced by the PMNS neutrino mixing matrix, $U_{PMNS}$ \cite{PMNS}. The standard parameterization of $U_{PMNS}$ with the CP-violating phases denoted by $\delta$, $\rho$ and $\sigma$ is given by
%%%%%%%
\begin{eqnarray}
U_{PMNS}=U_\nu K
\label{Eq:PMNS}
\end{eqnarray}
%%%%%%%
with
%%%%%%%
\begin{eqnarray}
U_\nu&=&\left( {\begin{array}{*{20}c}
   1 & 0 & 0  \\
   0 & {\cos \theta _{23} } & {\sin \theta _{23} }  \\
   0 & { - \sin \theta _{23} } & {\cos \theta _{23} } \\
\end{array}} \right)
\left( {\begin{array}{*{20}c}
   {\cos \theta _{13} } & 0 & {\sin \theta _{13}e^{-i\delta} }  \\
   0 & 1 & 0  \\
   { - \sin \theta _{13}e^{i\delta} } & 0 & {\cos \theta _{13} }  \\
\end{array}} \right)
\left( {\begin{array}{*{20}c}
   {\cos \theta _{12} } & {\sin \theta _{12} } & 0  \\
   { - \sin \theta _{12} } & {\cos \theta _{12} } & 0  \\
   0 & 0 & 1  \\
\end{array}} \right)
\nonumber \\
&=&\left( \begin{array}{ccc}
  c_{12}c_{13} &  s_{12}c_{13}&  s_{13}e^{-i\delta}\\
  -c_{23}s_{12}-s_{23}c_{12}s_{13}e^{i\delta}
                                 &  c_{23}c_{12}-s_{23}s_{12}s_{13}e^{i\delta}
                                 &  s_{23}c_{13}\\
  s_{23}s_{12}-c_{23}c_{12}s_{13}e^{i\delta}
                                 &  -s_{23}c_{12}-c_{23}s_{12}s_{13}e^{i\delta}
                                 & c_{23}c_{13}\\
\end{array} \right),
\label{Eq:U_nu}
\\
K &=& {\rm diag}(e^{i\rho}, e^{i\sigma}, 1),
\label{Eq:K}
\end{eqnarray}
%%%%%%%
where $c_{ij}=\cos\theta_{ij}$ and $s_{ij}=\sin\theta_{ij}$, if observed neutrino oscillations are induced by the mixing among massive Majorana neutrinos of the three flavors, $\nu_{e,\mu,\tau}$.  The recent experimental data \cite{RecentData,RecentAnalyses} show that 
%%%%%%%
\begin{eqnarray}
0.67\leq \sin^22\theta_\odot \leq 0.93, \quad \sin^22\theta_{atm} \geq 0.85, \quad \vert U_{e3}\vert^2 = \sin^2\theta_{13} < 0.048,
\label{Eq:Data}
\end{eqnarray}
%%%%%%%
where $\theta_\odot=\theta_{12}$ and $\theta_{atm}=\theta_{23}$, and that the magnitudes of three massive neutrino masses, $m_{1,2,3}$, are constrained as
%%%%%%%
\begin{eqnarray}
\Delta m^2_{atm} = \vert m^2_3-m^2_2\vert = (1.1-3.4)\times 10^{-3}~{\rm eV}^2,
\quad
\Delta m^2_\odot = \vert m^2_2-m^2_1\vert = (5.4-9.4)\times 10^{-5}~{\rm eV}^2.
\label{Eq:DataMass}
\end{eqnarray}
%%%%%%%

These experimental results create mysteries why the observed values are so realized in neutrino oscillations. The presence of tiny masses for neutrinos implied by Eq.(\ref{Eq:DataMass}) is understood by the seesaw mechanism \cite{Seesaw,type2seesaw} and by the radiative mechanism \cite{Zee,Babu}.  It is also stressed that the oscillations show 1) a hierarchy of $\Delta m^2_{atm} \gg \Delta m^2_\odot$ and 2) $\sin^22\theta_{atm,\odot}={\mathcal{O}}(1)$ while $\sin^2\theta_{13} \ll 1$.  Among various theoretical proposals to find clues behind the mystery, there are theoretical ideas based on 1) a conservation of the $L_e-L_\mu-L_\tau$ ($\equiv L^\prime$) number \cite{Lprime} and 2) a $\mu$-$\tau$ permutation symmetry \cite{Nishiura,MassMatrix,MaximalTexture,YasueMuTau,NewestMuTau}.  The hierarchy of $\Delta m^2_{atm} \gg \Delta m^2_\odot$ is due to the ideal situation with $\Delta m^2_\odot = 0$ and $\Delta m^2_{atm}\neq 0$, which are accounted by the $L^\prime$-conservation, while the maximal mixing of $\sin^2 2\theta_{atm} = 1$ arises from the $\mu$-$\tau$ permutation symmetry.  However, to discuss how to depart from these ideal cases is physically important.  Furthermore, if CP-violating phases are included, the possible form of the neutrino mass matrix is not fully understood \cite{MassTextureCP}.  We would like to discuss it to clarify its general properties and implications on neutrino physics. 

\section{\label{sec:2}Neutrino Mass Matrix}
The neutrino mass terms are described by 
%%%%%%%%%%%%%%%%%%%%
\begin{eqnarray}
-\mathcal{L}_{mass} &=& \frac{1}{2}
\left( {\nu _e ,\nu _\mu  ,\nu _\tau  } \right)^T M_\nu  \left( \begin{array}{l}
 \nu _e  \\ 
 \nu _\mu   \\ 
 \nu _\tau   \\ 
 \end{array} \right)
   	+ {\rm h.c.},
\label{Eq:NuMassTerm}
\end{eqnarray}
%%%%%%%%%%%%%%%%%%%%
where $M_\nu$ is a complex symmetric mass matrix.  It is understood that the charged leptons and neutrinos are rotated, if necessary, to give diagonal charged-current interactions and to define $\nu_e$, $\nu_\mu$ and $\nu_\tau$.  This flavor neutrino mass matrix can be diagonalized by $U_{PMNS}$ to give
%%%%%%%%%%%%%%%%%%%%
\begin{eqnarray}
U^T_{PMNS}M_\nu U_{PMNS} = M^{diag}_\nu = \left( {\begin{array}{*{20}c}
   {m_1 } & 0 & 0  \\
   0 & {m_2 } & 0  \\
   0 & 0 & {m_3 }  \\
\end{array}} \right).
\label{Eq:DiagonalMassTerm}
\end{eqnarray}
%%%%%%%%%%%%%%%%%%%%
Since $M_\nu$ is not Hermitian, one has to deal with the complexity due to the existence of all three phases, one Dirac phase of $\delta$ \cite{DiracPhase} and two Majorana phases of $\rho$ and $\sigma$ \cite{MajoranaPhase}. As a simpler choice, we use a Hermitian matrix \cite{MassSquredMatrix,MaxCPViolation}: 
%%%%%%%%%%%%%%%%%%%%
\begin{eqnarray}
&& M = M^\dagger_\nu M_\nu,
\nonumber \\
&& U^\dagger_{PMNS}M U_{PMNS} = \left( {\begin{array}{*{20}c}
   {m^2_1 } & 0 & 0  \\
   0 & {m^2_2 } & 0  \\
   0 & 0 & {m^2_3 }  \\
\end{array}} \right),
\label{Eq:HermitianMatrix}
\end{eqnarray}
%%%%%%%%%%%%%%%%%%%%
to examine the structure of $M_\nu$ so that two Majorana phases in $K$ become irrelevant. We parameterize $M_\nu$ by 
%%%%%%%%%%%%%%%%%%%%
\begin{eqnarray}
&& M_\nu = \left( {\begin{array}{*{20}c}
   a & b & c  \\
   b & d & e  \\
   c & e & f  \\
\end{array}} \right),
\label{Eq:NuMatrixEntries}
\end{eqnarray}
%%%%%%%%%%%%%%%%%%%%
leading to
%%%%%%%%%%%%%%%%%%%%
\begin{eqnarray}
&& M = \left( {\begin{array}{*{20}c}
   A & B & C  \\
   B^\ast & D & E  \\
   C^\ast & E^\ast & F  \\
\end{array}} \right),
\label{Eq:NuMatrixEntries2}
\end{eqnarray}
%%%%%%%%%%%%%%%%%%%%
where
%%%%%%%%%%%%%%%%%%%%
\begin{eqnarray}
&&
A = \vert a\vert^2 + \vert b\vert^2 + \vert c\vert^2, 
\quad
B = a^\ast b + b^\ast d + c^\ast e,
\quad
C = a^\ast c + b^\ast e + c^\ast f,
\nonumber \\
&&
D = \vert b\vert^2 + \vert d\vert^2 + \vert e\vert^2, 
\quad
E = b^\ast c + d^\ast e + e^\ast f,
\quad
F = \vert c\vert^2 + \vert e\vert^2 + \vert f\vert^2.
\label{Eq:EachNuMatrixEntries2}
\end{eqnarray}
%%%%%%%%%%%%%%%%%%%%
Note that $B$, $C$ and $E$ are complex. 

To examine the possible form of $M_\nu$ compatible with the observed data of Eqs(\ref{Eq:Data}) and (\ref{Eq:DataMass}), we have directly performed the computation of Eq.(\ref{Eq:HermitianMatrix}) and have found the following constraints:
%%%%%%%%%%%%%%%%%%%%
\begin{eqnarray}
&&
c_{12} \Delta_1  - s_{12} \left[ {\tilde s_{13}^{\ast}  \left( {c_{23} B^{\ast}   - s_{23} C^{\ast}  } \right) + c_{13} \Delta_2 } \right] = 0,
\quad
s_{12} \Delta_1  + c_{12} \left[ {\tilde s_{13}^{\ast}  \left( {c_{23} B^{\ast}   - s_{23} C^{\ast}  } \right) + c_{13} \Delta_2 } \right] = 0,
\label{Eq:Constraint1} \\
&&
c_{12} \left( {s_{12} \lambda_1  + c_{12} \left[ {c_{13} \left( {c_{23} B - s_{23} C} \right) - \tilde s_{13}^\ast  \Delta_2^{\ast} } \right]} \right) - s_{12} \left( c_{12} \lambda _2+{s_{12} \left[ {c_{13} \left( {c_{23} B^\ast   - s_{23} C^\ast  } \right) - \tilde s_{13} \Delta_2 } \right]} \right) = 0,
\label{Eq:Constraint2}
\end{eqnarray}
%%%%%%%%%%%%%%%%%%%%
for $\tilde s_{13}=s_{13}e^{i\delta}$, where $\Delta_{1,2}$ are defined to be: 
%%%%%%%%%%%%%%%%%%%%
\begin{eqnarray}
&&
\Delta_1  = c_{13} \tilde s_{13}^{\ast}  \left( {A - \lambda _3 } \right) + c_{13}^2 \left( {s_{23} B + c_{23} C} \right) - \tilde s_{13}^{{\ast}2} \left( {s_{23} B^{\ast}   + c_{23} C^{\ast} } \right),
\quad
\Delta_2  = c_{23}^2 E - s_{23}^2 E^{\ast}   + s_{23} c_{23} \left( {D - F} \right), 
\label{Eq:Constraint3}
\end{eqnarray}
%%%%%%%%%%%%%%%%%%%%
and the diagonalized masses:
%%%%%%%%%%%%%%%%%%%%
\begin{eqnarray}
&&
m_1^2  = c_{12}^2 \lambda_1  + s_{12}^2 \lambda _2  - 2c_{12} s_{12}X,
\quad
m_2^2  = s_{12}^2 \lambda_1  + c_{12}^2 \lambda _2  + 2c_{12} s_{12}X,
\nonumber \\
&&
m_3^2  =c_{13}^2 \lambda _3 + s_{13}^2 A + c_{13} \left[ {\tilde s_{13} \left( {s_{23} B + c_{23} C} \right) + \tilde s_{13}^{\ast}  \left( {s_{23} B^{\ast}   + c_{23} C^{\ast}  } \right)} \right],
\label{Eq:Masses}
\end{eqnarray}
%%%%%%%%%%%%%%%%%%%%
and
%%%%%%%%%%%%%%%%%%%%
\begin{eqnarray}
&&
\lambda_1  = c_{13}^2 A - c_{13} \left[ {\tilde s_{13} \left( {s_{23} B + c_{23} C} \right) + \tilde s_{13}^{\ast}  \left( {s_{23} B^{\ast}   + c_{23} C^{\ast}  } \right)} \right] + s_{13}^2 \lambda _3, 
\nonumber \\
&&
\lambda _2  = c_{23}^2 D + s_{23}^2 F - 2s_{23} c_{23} {\rm Re}\left( E\right), 
\quad
\lambda _3  = s_{23}^2 D + c_{23}^2 F + s_{23} c_{23} {\rm Re}\left( E\right).
\nonumber \\
&&
2X=c_{13} \left( {c_{23} B - s_{23} C} \right) - \tilde s_{13}^{\ast}  \Delta_2 ^{\ast}   + c_{13} \left( {c_{23} B^{\ast}   - s_{23} C^{\ast}  } \right) - \tilde s_{13} \Delta_2,
\label{Eq:Parameters}
\end{eqnarray}
%%%%%%%%%%%%%%%%%%%%
Since Eq.(\ref{Eq:Constraint1}) gives
%%%%%%%%%%%%%%%%%%%%
\begin{eqnarray}
&&
\Delta_1=0,
\quad
\tilde s_{13}^{\ast}  \left( {c_{23} B^{\ast} - s_{23} C^{\ast}  } \right) + c_{13} \Delta_2 = 0,
\label{Eq:Delta_Deltalmbda}
\end{eqnarray}
%%%%%%%%%%%%%%%%%%%%
we obtain that
%%%%%%%%%%%%%%%%%%%%
\begin{eqnarray}
&&
\tan 2\theta_{13}  = 2\frac{{\left| {s_{23} B + c_{23} C} \right|}}{{\lambda _3  - A}},
\quad
X=\frac{c_{23} {\rm Re} \left( B \right) - s_{23} {\rm Re} \left( C \right)}{c_{13}},
\label{Eq:theta13}
\end{eqnarray}
%%%%%%%%%%%%%%%%%%%%
and $\delta$ used in $\tilde s_{13}$ is determined by the phase of $s_{23} B + c_{23} C$ to be:
%%%%%%%%%%%%%%%%%%%%
\begin{eqnarray}
s_{23} B + c_{23} C = \left| {s_{23} B + c_{23} C} \right| e^{-i\delta },
\label{Eq:Phase-delta}
\end{eqnarray}
%%%%%%%%%%%%%%%%%%%%
provided that $s_{23}B+c_{23}C\neq 0$.  By using Eq.(\ref{Eq:Constraint2}) with Eq.(\ref{Eq:Delta_Deltalmbda}) for $\Delta_2$, we find that
%%%%%%%%%%%%%%%%%%%%
\begin{eqnarray}
&&
\tan \theta _{23}  = \frac{{\rm Im}\left( B\right)}{{\rm Im}\left( C\right)},
\label{Eq:theta23}
\end{eqnarray}
%%%%%%%%%%%%%%%%%%%%
from the constraint on the imaginary part:
%%%%%%%%%%%%%%%%%%%%
\begin{eqnarray}
&&
c_{23}B-s_{23}C=c_{23}B^\ast-s_{23}C^\ast,
\label{Eq:RealCond}
\end{eqnarray}
%%%%%%%%%%%%%%%%%%%%
and
%%%%%%%%%%%%%%%%%%%%
\begin{eqnarray}
&&
\tan 2\theta _{12}  = 2\frac{X}{\lambda _2  - \lambda_1},
\label{Eq:theta12}
\end{eqnarray}
%%%%%%%%%%%%%%%%%%%%
from the constraint on the real part.  As a result, $\delta$ is expressed as:
%%%%%%%%%%%%%%%%%%%%
\begin{eqnarray}
&&
\tan \delta  = -\frac{1}{{s_{23} }}\frac{{{\rm Im} \left( B \right)}}{{s_{23} {\rm Re} \left( B \right) + c_{23} {\rm Re} \left( C \right)}}.
\label{Eq:thetaAlpha}
\end{eqnarray}
%%%%%%%%%%%%%%%%%%%%
Furthermore, considering the relations of $\Delta_2$ in Eqs.(\ref{Eq:Constraint3}) and (\ref{Eq:Delta_Deltalmbda}), we also obtain that
%%%%%%%%%%%%%%%%%%%%
\begin{eqnarray}
&&
{\rm Im} \left( E \right) =  s_{13} X \sin \delta, 
\nonumber \\ 
&&
\cos 2\theta _{23} {\rm Re} \left( E \right) = \frac{{\sin 2\theta _{23} }}{2}\left( {F - D} \right) - s_{13} X\cos \delta. 
\label{Eq:OrherConstraint}
\end{eqnarray}
%%%%%%%%%%%%%%%%%%%%
From these relations, the mass parameters are further converted to give
%%%%%%%%%%%%%%%%%%%%
\begin{eqnarray}
&&
m^2_1 = \frac{\lambda_1+\lambda_2}{2}-\frac{X}{\sin 2\theta_{12}},
\quad
m^2_2 = \frac{\lambda_1+\lambda_2}{2}+\frac{X}{\sin 2\theta_{12}},
\quad
m^2_3 = \frac{c^2_{13}\lambda_3-s^2_{13}A}{c^2_{13}-s^2_{13}},
\label{Eq:masses2_1-2-3}
\end{eqnarray}
%%%%%%%%%%%%%%%%%%%%
where
%%%%%%%%%%%%%%%%%%%%
\begin{eqnarray}
&&
\lambda_1 = \frac{c^2_{13}A-s^2_{13}\lambda_3}{c^2_{13}-s^2_{13}},
\label{Eq:lambda_1}
\end{eqnarray}
%%%%%%%%%%%%%%%%%%%%
together with Eq.(\ref{Eq:Parameters}) for $\lambda_{2,3}$.  We find that
%%%%%%%%%%%%%%%%%%%%
\begin{eqnarray}
&&
\Delta m^2_\odot = 2\frac{c_{23} {\rm Re} \left( B \right) - s_{23} {\rm Re} \left( C \right)}{c_{13}\sin 2\theta_{12}},
\label{Eq:Solar_m2}
\end{eqnarray}
%%%%%%%%%%%%%%%%%%%%
which is the useful relation.

If other parameterizations of the CP-violating Dirac phase in $U_{PMNS}$ are employed, different relations will be derived.  However, this difference can be absorbed in the redefinition of the masses.  For example, results from $U_{PMNS}$ of the Kobayashi-Maskawa type (with $e^{-i\delta}$ as a CP-violating phase) \cite{K-M} can be generated by the replacement of $B$ and $C$ by $Be^{-i\delta}$ and $Ce^{-i\delta}$ in the same $M$ as Eq.(\ref{Eq:NuMatrixEntries2}) with the standard $U_{PMNS}$ of Eq.(\ref{Eq:U_nu}).  Especially, in Eq.(\ref{Eq:Phase-delta}), $s_{23} B + c_{23} C = \left| {s_{23} B + c_{23} C} \right| e^{-i\delta }$ becomes $s_{23} Be^{-i\delta } + c_{23} Ce^{-i\delta } = \left| {s_{23} B + c_{23} C} \right| e^{-i\delta }$, leading to $s_{23} B + c_{23} C = \left| {s_{23} B + c_{23} C} \right|$(=real). Therefore, no CP-violating phase is induced by Eq.(\ref{Eq:Phase-delta}). Instead, the CP-violating phase is induced by $c_{23}B-s_{23}C=\vert c_{23}B-s_{23}C\vert e^{i\delta}$ derived as a solution of $(s_{23}B-c_{23}C)e^{-i\delta}$=$(s_{23}B^\ast-c_{23}C^\ast)e^{i\delta}$ from Eq.(\ref{Eq:RealCond}) with the appropriate replacement of $B$ and $C$.  Since $s_{23} B + c_{23} C$=real, the relation of $\tan \theta_{23}$ becomes $\tan \theta_{23} = -{\rm Im}(C)/{\rm Im}(B)$ instead of Eq.(\ref{Eq:theta23}).  In this article, we only show the results by using the standard $U_{PMNS}$.

Since $\sin^2\theta_{13} < 0.048$ is reported, let us choose $\sin\theta_{13}=0$ and no CP-violation phase is induced by Eq.(\ref{Eq:Phase-delta}) because of $s_{23} B + c_{23} C = 0$ in Eq.(\ref{Eq:theta13}), leading to real $B$ and $C$ from Eq.(\ref{Eq:RealCond}). The mixing angle of $\theta_{23}$ is determined by \cite{YasueMuTau}
%%%%%%%%%%%%%%%%%%%%
\begin{eqnarray}
&&
\tan\theta_{23} = -\frac{C}{B},
\label{Eq:theta_12_real}
\end{eqnarray}
%%%%%%%%%%%%%%%%%%%%
which should be compared with Eq.(\ref{Eq:theta23}) for the complex $B$ and $C$.  We obtain that ${\rm Im} (E) =  0$ and
%%%%%%%%%%%%%%%%%%%%
\begin{eqnarray}
&&
2E\cos 2\theta_{23} = \left(F-D\right) \sin 2\theta_{23},
\label{Eq:Solution_13=0}
\end{eqnarray}
%%%%%%%%%%%%%%%%%%%%
from Eq.(\ref{Eq:OrherConstraint}) with $t_{13}=0$.  The solar neutrino mixing angle of $\theta_{12}$ is given by
%%%%%%%%%%%%%%%%%%%%
\begin{eqnarray}
&&
\tan 2\theta_{12} = \frac{2B}{c_{23}\left( \lambda_2 - \lambda_1\right)}.
\label{Eq:Solution_12}
\end{eqnarray}
%%%%%%%%%%%%%%%%%%%%

The observed atmospheric neutrino mixing is close to the maximal one.  We, then, restrict ourselves to the case with $\tan\theta_{23}=\pm 1~(\equiv \kappa)$ and $\cos\theta_{23}>0$.  The relation in Eq.(\ref{Eq:theta23}) becomes
%%%%%%%%%%%%%%%%%%%%
\begin{eqnarray}
&&
{\rm Im}\left( C\right) = \kappa {\rm Im}\left(B\right),
\label{Eq:Maximal}
\end{eqnarray}
%%%%%%%%%%%%%%%%%%%%
and  the constranint on $E$ becomes
%%%%%%%%%%%%%%%%%%%%
\begin{eqnarray}
&&
\kappa \left( F - D\right) = \sqrt{2}t_{13} \cos \delta\left( {\rm Re} \left( B \right) - \kappa{\rm Re} \left( C \right) \right),
\quad
\sqrt{2}{\rm Im} \left( E \right) =   t_{13} \sin \delta\left( {\rm Re} \left( B \right) - \kappa{\rm Re} \left( C \right) \right). 
\label{Eq:OrherMaximalConstraint}
\end{eqnarray}
%%%%%%%%%%%%%%%%%%%%
If no CP-violation exists, $\delta=0$ is required.  Namely, it demands ${\rm Im}(B)=0$ from Eq.(\ref{Eq:thetaAlpha}), which, in turn, gives $\tan\theta_{23}=0$ if ${\rm Im}(C)\neq 0$. To get around $\tan\theta_{23}=0$, ${\rm Im}(C)=0$ should be imposed and Eq.(\ref{Eq:Maximal}) disappears.  As a result, $B$, $C$ and $E$ turn out to be all real. The mixing angle of $\theta_{23}$ is determined by $\kappa$ derived from Eq.(\ref{Eq:OrherMaximalConstraint}) with $\delta=0$. Therefore, in order to ensure the appearance of the maximal atmospheric neutrino mixing, we have to be careful to impose the constraint:
%%%%%%%%%%%%%%%%%%%%
\begin{itemize}
\item ${\rm Im}\left( C\right) = \kappa{\rm Im}\left( B\right)$ for the presence of CP-violation,
\item $\kappa\left( F - D\right) = \sqrt{2}t_{13}(B - \kappa C)$ with real $B$ and $C$ for the absence of CP-violation. 
\end{itemize}
%%%%%%%%%%%%%%%%%%%%
It should be noted that, for the absence of CP-violation, the approximate equality of $F\sim D$ is often assumed and it results in the familiar relation of $\tan\theta_{23}(=\kappa) \sim -C/B$ for $\vert t_{13}\vert \sim 0$. Another solution with $\tan\theta_{23}(=\kappa) \sim B/C$ seems appropriate because it may yield $\sin^22\theta_{12}\sim 0$ in Eq.(\ref{Eq:theta12}).

\section{\label{sec:3}Examples}
To see how the present analysis works, let us examine the specific example of $M_\nu$ with $c=-\kappa b^\ast$ given by
%%%%%%%%%%%%%%%%%%%%
\begin{eqnarray}
&& 
	M_\nu = \left( {\begin{array}{*{20}c}
   a & b & -\kappa b^\ast  \\
   b & d & e  \\
   -\kappa b^\ast & e & d^\ast  \\
\end{array}} \right),
\label{Eq:SpecificMnu}
\end{eqnarray}
%%%%%%%%%%%%%%%%%%%%
whose physical consequence has been discussed in Ref.\cite{SpecificMnu}.  From $M^\dagger_\nu M_\nu$, we obtain that
%%%%%%%%%%%%%%%%%%%%
\begin{eqnarray}
&&
A = \vert a\vert^2 + 2\vert b\vert^2, 
\quad
B = \left( a^\ast -\kappa e\right) b + b^\ast d,
\quad
C = \left(-\kappa  a^\ast + e\right) b^\ast -\kappa b d^\ast,
\nonumber \\
&&
D = F = \vert b\vert^2 + \vert d\vert^2 + \vert e\vert^2, 
\quad
E = -\kappa b^\ast b^ \ast + 2d^\ast{\rm Re}\left( e\right).
\label{Eq:SpecificMnu2-1}
\end{eqnarray}
%%%%%%%%%%%%%%%%%%%%
To meet the maximal atmospheric neutrino mixing, one simple choice of mass terms is to assume that $a$ and $e$ are real so that $C=-\kappa B^\ast$, leading to ${\rm Re}(C)=-\kappa {\rm Re}(B)$ and ${\rm Im}(C)=\kappa {\rm Im}(B)$. Therefore, $\tan\theta_{23}=\kappa$ is recovered. From Eq.(\ref{Eq:thetaAlpha}), we have 
%%%%%%%%%%%%%%%%%%%%
\begin{eqnarray}
&&
\tan \delta  = -\frac{{2{\rm Im} \left( B \right)}}{{\rm Re} \left( B \right) +\kappa{\rm Re} \left( C \right)}=\pm\infty,
\label{Eq:tan_alpha}
\end{eqnarray}
%%%%%%%%%%%%%%%%%%%%
leading to $\vert\delta\vert = \pi/2$, which indicates the maximal CP-violation \cite{SpecificMnu}.  From Eq.(\ref{Eq:OrherMaximalConstraint}), we also have $F=D$ by $\cos\delta=0$ consistent with Eq.(\ref{Eq:SpecificMnu2-1}) as expected and ${\rm Im}(E)=\pm\sqrt{2}\vert t_{13}{\rm Re}(B)\vert$ as an additional constraint, where $\pm$ depends on the sign of $\delta$.  Other mixing angles are computed to be:
%%%%%%%%%%%%%%%%%%%%
\begin{eqnarray}
&&
\tan 2\theta_{12}  = 2\sqrt{2} \frac{{\rm Re} \left( B \right)}{c_{13}\left( D-\kappa {\rm Re}(E)-\lambda_1 \right)},
\quad
\tan 2\theta_{13}  = 2\sqrt{2}\frac{{\rm Im} \left( B \right)}{D+\kappa {\rm Re}(E)-A}.
\label{Eq:tan_alpha}
\end{eqnarray}
%%%%%%%%%%%%%%%%%%%%

Another example is $M_\nu$ with a $\mu$-$\tau$ permutation symmetry \cite{Nishiura,MassMatrix,MaximalTexture,YasueMuTau,NewestMuTau}, which suggests that $c = \kappa b$ and $d=f$ giving
%%%%%%%%%%%%%%%%%%%%
\begin{eqnarray}
&& 
	M_\nu = \left( {\begin{array}{*{20}c}
   a & b & \kappa b  \\
   b & d & e  \\
   \kappa b & e & d  \\
\end{array}} \right).
\label{Eq:SpecificMnu2-2}
\end{eqnarray}
%%%%%%%%%%%%%%%%%%%%
From this matrix,
%%%%%%%%%%%%%%%%%%%%
\begin{eqnarray}
&&
A = \vert a\vert^2 + 2\vert b\vert^2, 
\quad
B = a^\ast b + b^\ast d + \kappa b^\ast e,
\quad
C = \kappa a^\ast b + b^\ast e + \kappa b^\ast d,
\nonumber \\
&&
D = F = \vert b\vert^2 + \vert d\vert^2 + \vert e\vert^2, 
\quad
E = \kappa \vert b\vert^2 + d^\ast e + e^\ast d,
\label{Eq:mu-tau2}
\end{eqnarray}
%%%%%%%%%%%%%%%%%%%%
are obtained.  Since $C=\kappa B$ and $X=0$, we find that $\tan\theta_{23}=\kappa$ and $\tan 2\theta_{12} = 0$ unless $\lambda_1 = \lambda_2$.  If $\lambda_1 = \lambda_2$, $\theta_{12}$ is not fixed and the masses of $m_{1,2}$ turn out to satisfy $m_1^2  = m_2^2$ and $\Delta m^2_\odot = 0$. The mixing angle of $\theta_{13}$ and the CP-violating phase of $\delta$ are determined to be:
%%%%%%%%%%%%%%%%%%%%
\begin{eqnarray}
&&
\tan 2\theta_{13} = 2\sqrt{2}\frac{\vert B\vert}{D+\kappa E-A},
\quad
\tan\delta =-\frac{{\rm Im}\left( B \right)}{{\rm Re}\left( B \right)}.
\label{Eq:theta13_alpha}
\end{eqnarray}
%%%%%%%%%%%%%%%%%%%%
Since $\sin^2 2\theta_{12}=4/(4+x^2)$ with $x = (\lambda_2-\lambda_1)/X$, we may require that $x\sim 1$:
%%%%%%%%%%%%%%%%%%%%
\begin{eqnarray}
\lambda_2 - \lambda_1 \sim  X(=( B - \kappa C)/c_{13}),
\label{Eq:theta13_alpha}
\end{eqnarray}
%%%%%%%%%%%%%%%%%%%%
that gives $\sin^2 2\theta_{12}\sim 0.8$ after the $\mu$-$\tau$ symmetry is slightly at least broken by ${\rm Re}(B)\neq {\rm Re}(C)$, which also induces $\Delta m^2_\odot \neq 0$.

Finally, let us examine a typical texture among mass matrices with two texture zeros \cite{TwoTextureZeros}, which is given by
%%%%%%%%%%%%%%%%%%%%
\begin{eqnarray}
&& 
	M_\nu = \left( {\begin{array}{*{20}c}
   0 & b & 0 \\
   b & d & e(=\kappa d)  \\
   0 & e(=\kappa d) & f  \\
\end{array}} \right),
\label{Eq:SpecificMnu2-3}
\end{eqnarray}
%%%%%%%%%%%%%%%%%%%%
where the similar matrix with $b=0$ and $c\neq 0$ can be treated in the same way, and it yields
%%%%%%%%%%%%%%%%%%%%
\begin{eqnarray}
&&
A = \vert b\vert^2, 
\quad
B = b^\ast d,
\quad
C = b^\ast e,
\nonumber \\
&&
D = \vert b\vert^2 + \vert d\vert^2 + \vert e\vert^2, 
\quad
E = d^\ast e + e^\ast f,
\quad
F = \vert e\vert^2 + \vert f\vert^2.
\label{Eq:mu-tau2}
\end{eqnarray}
%%%%%%%%%%%%%%%%%%%%
The simplest way to get the maximal atmospheric neutrino mixing is to require that $e=\kappa d$, leading to $C=\kappa B$ and $\tan\theta_{23}=\kappa$.  Then, the same relations for $\tan\theta_{13}$ and $\tan\delta$ as those in the $\mu$-$\tau$ symmetric case are also satisfied and, accordingly, $\theta_{12}$ is left undetermined.  Additional constraints are given by ${\rm arg}(d)={\rm arg}(f)$ from ${\rm Im}(E)=0$ and $\vert f\vert^2 = \vert b\vert^2 + \vert d\vert^2$ from $F=D$ imposed by Eq.(\ref{Eq:OrherMaximalConstraint}).

\section{\label{sec:4}Discussions}
In summary, if CP-violation is present, using the complex symmetric mass matrix given by
%%%%%%%%%%%%%%%%%%%%
\begin{eqnarray}
&& M_\nu = \left( {\begin{array}{*{20}c}
   a & b & c  \\
   b & d & e  \\
   c & e & f  \\
\end{array}} \right),
\label{Eq:NuMatrixEntries_summary}
\end{eqnarray}
%%%%%%%%%%%%%%%%%%%%
we have found the important and simple relation:
%%%%%%%%%%%%%%%%%%%%
\begin{eqnarray}
&&
\tan \theta _{23}  = \frac{{\rm Im}\left( B\right)}{{\rm Im}\left( C\right)},
\label{Eq:theta_23_summary}
\end{eqnarray}
%%%%%%%%%%%%%%%%%%%%
for $B = a^\ast b + b^\ast d + c^\ast e$ and $C = a^\ast c + b^\ast e + c^\ast f$.  Any models with CP-violation should respect $\vert {\rm Im}\left( B\right) \vert \sim \vert {\rm Im}\left( C\right) \vert$ to explain the observed result of $\sin^22\theta_{23} \sim 1$.  Furthermore, the CP-violating Dirac phase is induced by the phase of $s_{23} B + c_{23} C$. The physically interesting quantity of $\theta_{13}$ is determined to be:
%%%%%%%%%%%%%%%%%%%%
\begin{eqnarray}
&&
\tan 2\theta_{13}  = \frac{2{\left| {s_{23} B + c_{23} C} \right|}}{{s_{23}^2 D + c_{23}^2 F + 2s_{23} c_{23}{\rm Re}\left( E\right)  - A}},
\label{Eq:theta_13_summary}
\end{eqnarray}
%%%%%%%%%%%%%%%%%%%%
for $A = \vert a\vert^2 + \vert b\vert^2 + \vert c\vert^2$, $D = \vert b\vert^2 + \vert d\vert^2 + \vert e\vert^2$, $E = b^\ast c + d^\ast e + e^\ast f$ and $F = \vert c\vert^2 + \vert e\vert^2 + \vert f\vert^2$.  Other results include
%%%%%%%%%%%%%%%%%%%%
\begin{eqnarray}
&&
\Delta m^2_\odot = \frac{2X}{\sin 2\theta_{12}},
\quad
\tan 2\theta _{12}  = \frac{2X}{\lambda _2  - \lambda_1},
\label{Eq:Solar_m2_summary}
\end{eqnarray}
%%%%%%%%%%%%%%%%%%%%
where
%%%%%%%%%%%%%%%%%%%%
\begin{eqnarray}
X = \frac{c_{23} {\rm Re} \left( B \right) - s_{23} {\rm Re} \left( C \right)}{\cos\theta_{13}}.
\label{Eq:Solar_m2_summary}
\end{eqnarray}
%%%%%%%%%%%%%%%%%%%%

In order to understand the origin of $\sin^2\theta_{13}\ll 1$, it is instructive to notice that the mass terms are grouped into three categories according to the electron number $L_e$ \cite{eNumber,LeMass}. Namely, $a$ has $L_e=2$, $b$ and $c$ have $L_e=1$ and $d$, $e$ and $f$ have $L_e=0$.  If the mass terms with $L_e\neq 0$ are created by perturbative interactions with $\vert L_e\vert = 1$, we may assume that $\vert a\vert \ll \vert b,c\vert \ll \vert d,e,f\vert$, leading to $\vert A\vert \ll \vert B,C\vert \ll \vert D,E,F\vert$.  Then, this hierarchy ensures the appearance of $\vert \tan 2\theta_{13}\vert\ll 1$ due to the tiny violation of the $L_e$-conservation and at the same time may allow $\Delta m^2_{atm}\gg\Delta m^2_\odot$ to arise if $\Delta m^2_{atm} \mapgeq {\mathcal{O}}(|D,E,F|)$ since $\Delta m^2_\odot = {\mathcal{O}}(|{\rm Re}(B,C)|)$ \cite{LeMass}.

The smallest $\vert \tan 2\theta_{13}\vert$ can be acheived at $\tan\theta_{23}=-{\rm Re}(C)/{\rm Re}(B)$ to yield $\vert{\rm Im}(B)/s_{23}\vert$, pointing to $\tan\delta = \infty$ from Eq.(\ref{Eq:thetaAlpha}) and showing the maximal CP-violation. Considering $\tan \theta _{23} = {\rm Im}(B)/{\rm Im}(C)$, we observe that $B$ and $C$ are consistently related to be, for the maximal atmospheric neutrino mixing with $\tan\theta_{23}=\kappa(=\pm 1)$,
%%%%%%%%%%%%%%%%%%%%
\begin{eqnarray}
&&
C=- \kappa B^\ast,
\label{Eq:ConsistentRelation}
\end{eqnarray}
%%%%%%%%%%%%%%%%%%%%
which is the case of Eq.(\ref{Eq:SpecificMnu}) and also corresponds to the mass matrix in Ref.\cite{MaxCPViolation}.  The same result of the maximal CP-violation is obtained for $U_{PMNS}$ of the Kobayashi-Maskawa type if $\tan\theta_{23}={\rm Re}(B)/{\rm Re}(C)$ is chosen and $C=\kappa B^\ast$ becomes a consistent relation.

\vspace{3mm}

\noindent
\centerline{\bf Acknowledgements}

The authors would like to thank T. Kitabayashi for enlightening discussions.
%%-------------------------------------------------
%% References
%%-------------------------------------------------

\end{document}